\documentstyle[12pt,aaspp4,psfig,natbib]{article}

\newcommand{\vy}[2]{#1_{\scriptscriptstyle #2}}
\newcommand{\Ly}{Ly$\alpha$}

\def\gtorder{\mathrel{\raise.3ex\hbox{$>$}\mkern-14mu
             \lower0.6ex\hbox{$\sim$}}}
\def\ltorder{\mathrel{\raise.3ex\hbox{$<$}\mkern-14mu
             \lower0.6ex\hbox{$\sim$}}}
\def\proptwid{\mathrel{\raise.3ex\hbox{$\propto$}\mkern-14mu
             \lower0.6ex\hbox{$\sim$}}}
\def\arcsec{\ifmmode '' \else $''$\fi}

\def\arcsecpoint{\ifmmode ''\!. \else $''\!.$\fi}

\def\kms{\ifmmode {\rm km\ s}^{-1} \else km s$^{-1}$\fi}
\def\Msun{\ifmmode {\rm M}_{\odot} \else M$_{\odot}$\fi}
\def\Lsun{\ifmmode {\rm L}_{\odot} \else L$_{\odot}$\fi}
\def\Zsun{\ifmmode {\rm Z}_{\odot} \else Z$_{\odot}$\fi}

\def\ergscm2{ergs\,s$^{-1}$\,cm$^{-2}$}
\def\icm3{{\rm cm}^{-3}}
\def\icm2{{\rm cm}^{-2}}
\def\qo{\ifmmode q_{\rm o} \else $q_{\rm o}$\fi}
\def\Ho{\ifmmode H_{\rm o} \else $H_{\rm o}$\fi}
\def\ho{\ifmmode h_{\rm o} \else $h_{\rm o}$\fi}

\def\vFWHM{\ifmmode v_{\mbox{\tiny FWHM}} \else
            $v_{\mbox{\tiny FWHM}}$\fi}
\def\CCF{\ifmmode F_{\it CCF} \else $F_{\it CCF}$\fi}
\def\ACF{\ifmmode F_{\it ACF} \else $F_{\it ACF}$\fi}
\def\Halpha{\ifmmode {\rm H}\alpha \else H$\alpha$\fi}
\def\Hbeta{\ifmmode {\rm H}\beta \else H$\beta$\fi}
\def\Hgamma{\ifmmode {\rm H}\gamma \else H$\gamma$\fi}
\def\Hdelta{\ifmmode {\rm H}\delta \else H$\delta$\fi}
\def\Lya{\ifmmode {\rm Ly}\alpha \else Ly$\alpha$\fi}
\def\Lyb{\ifmmode {\rm Ly}\beta \else Ly$\beta$\fi}
\def\Lyg{\ifmmode {\rm Ly}\beta \else Ly$\gamma$\fi}
\def\hi{H\,{\sc i}}

\def\hei{He\,{\sc i}}

\def\cii{C\,{\sc ii}}
\def\ciii{\ifmmode {\rm C}\,{\sc iii} \else C\,{\sc iii}\fi}
\def\civ{\ifmmode {\rm C}\,{\sc iv} \else C\,{\sc iv}\fi}

\def\nv{N\,{\sc v}}

\def\o5007{[O\,{\sc iii}]\,$\lambda5007$}

\def\ovi{O\,{\sc vi}}

\def\mgii{Mg\,{\sc ii}}

\def\siiv{Si\,{\sc iv}}

\def\caii{Ca\,{\sc ii}}

\def\feii{Fe\,{\sc ii}}

\def\aliii{Al\,{\sc iii}}

\def\o{\o}
\begin{document}

\title{PHOTOIONIZATION MODELS FOR BALQSO PG 0946+301}


\author{
Nahum Arav\altaffilmark{1,2}, 
Kirk T. Korista\altaffilmark{3},
Martijn de~Kool\altaffilmark{4} 
}

\altaffiltext{1}{Astronomy Department, UC Berkeley, Berkeley, 
CA 94720, I:arav@astron.Berkeley.EDU}
\altaffiltext{2}{Physics Department, University of California, Davis, CA 95616}
\altaffiltext{3}{Western Michigan Univ.,Dept. of Physics,1120 Everett Tower, 
Kalamazoo, MI  49008}
\altaffiltext{4}{Research School of Astronomy and Astrophysics, ANU ACT,
 Australia}

\begin{abstract}

This is a companion paper to the manuscript entitled: ``HST STIS
Observations Of PG 0946+301: The Highest Quality UV Spectrum Of A
BALQSO'' (Arav et al. 2001; accepted for publication in the ApJ). Here
we present photoionization-modeling results for {\it all} the ionic
column density constraints found in these data, most of which we
were unable to include in the printed version of the paper.

\end{abstract}

\section{DESCRIPTION}

\begin{figure}
\centerline{\psfig{file=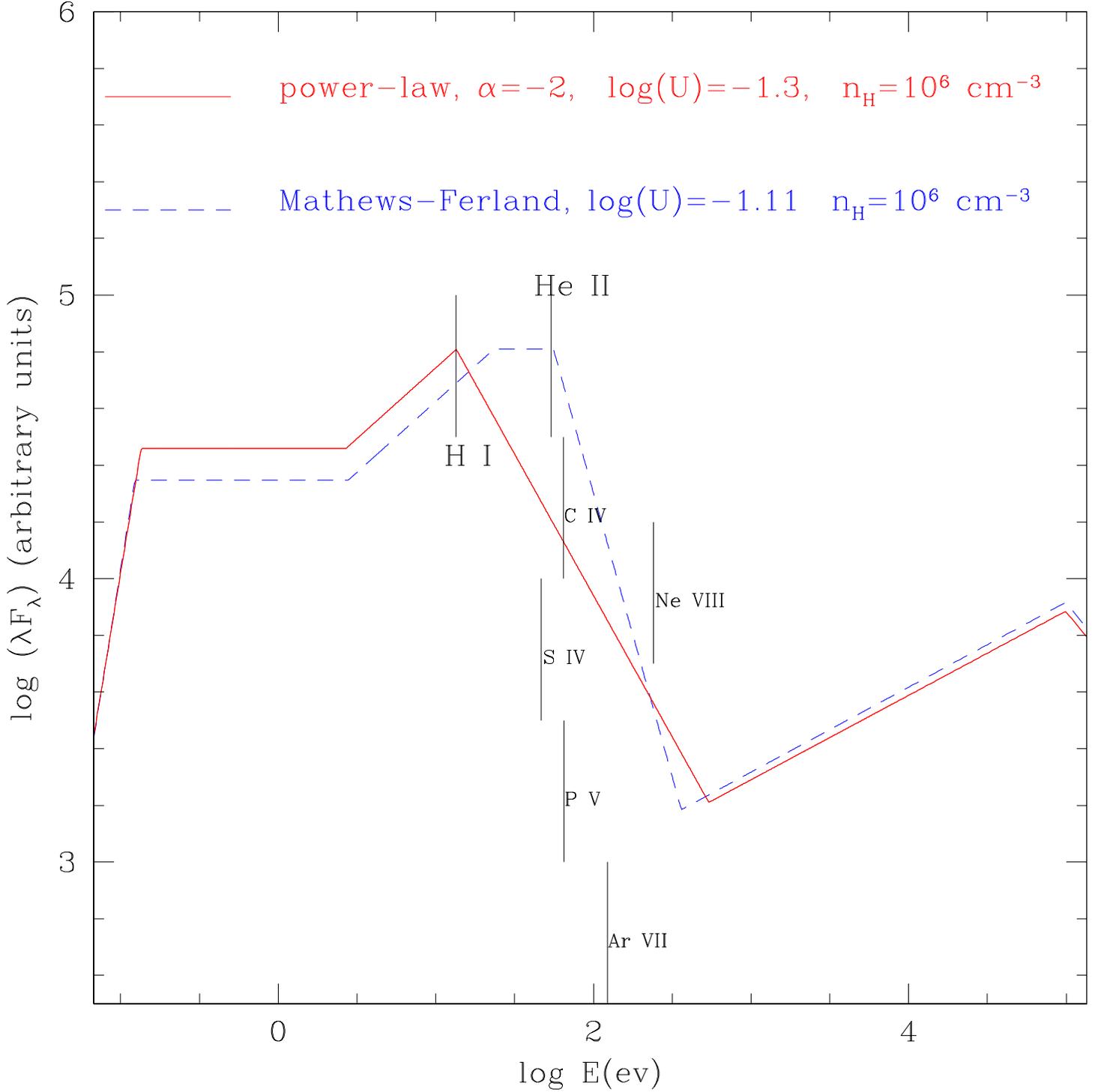,height=21.0cm,width=21.0cm}}
\caption{Comparison between Mathews--Ferland and a modified $\alpha=-2$ power-law spectrum,
which is consistent with the observed far UV spectral-shape of the PG 0946+301 spectrum.
Ionization potentials for several important ions are marked by short vertical lines.}
\label{mf_pl}
\end{figure}

In the following pages we present photoionization-modeling results
for {\it all} the ionic column densities constraints found in the
HST/STIS spectra of PG 0946+301.  We use the $\vy{N}{H}$/$U$ plane
presentation, applied to optically thin slabs, which is described in
\S~4 of the main paper.  Two incident spectra are used for the CLOUDY
models, the standard Mathews--Ferland  AGN spectrum and a modified
$\alpha$=--2 power-law spectrum which is more consistent with the
observed far UV spectral shape of the PG 0946+301 spectrum. Both
spectra are shown in figure \ref{mf_pl}.  Figures 2--4 show the
results for the modified $\alpha$=--2 power-law spectrum. In figure 2
we include only the hydrogen and helium results, for helium we have an
upper limit which therefore excludes the parameter area above the
curve.  Figure 3 shows results for the available ionic constraints from
8 ``metals.''  Each curve is labeled by its corresponding ion and the
number beneath the label is the $\log(N_{ion})$ constraint.  An *
following the $\log(N_{ion})$ denotes an upper limit (excluding the
parameter area above the curve), otherwise the constraint is a lower
limit (excluding the parameter area below the curve).  The small box
at the center of the plots identifies the area in parameter space
which is fully consistent with 21 ionic constraints, marginally
inconsistent with 3 and is strongly inconsistent with 2 (see \S~4 for
details).  Figure 4 is identical to figure 7 in the main paper.

Figures 5--8 are similar to figures 2--4, using the  
Mathews--Ferland  AGN spectrum. To ease the comparison between results from the 
two incident continua, we left the parameter space box at the same location
as for the power-law spectrum.

\clearpage

\begin{figure}
\centerline{\psfig{file=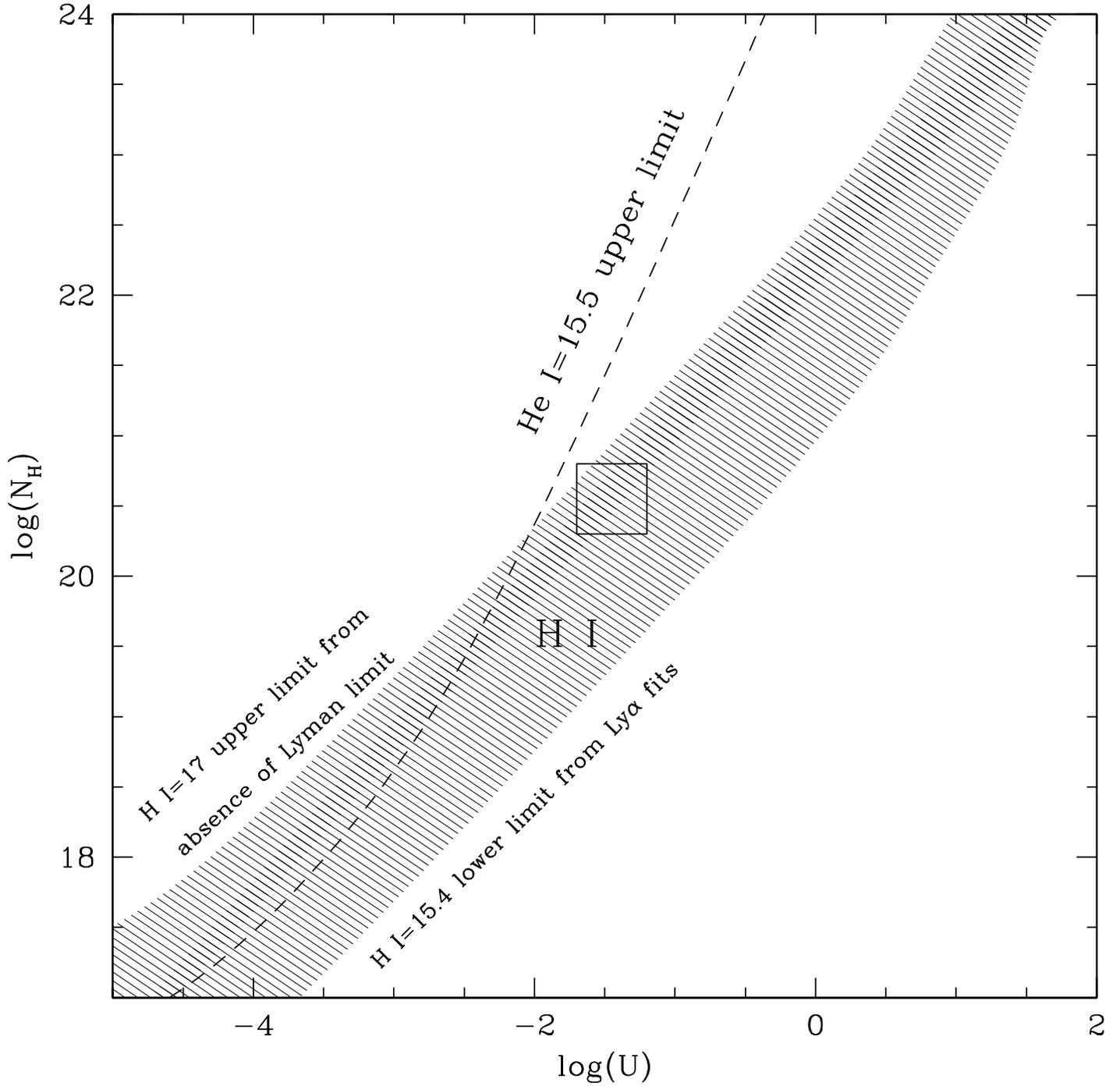,height=20.0cm,width=20.0cm}}
\caption{The all-important hydrogen constraints and their origin for the 
modified $\alpha=-2$ power-law spectrum.  Also shown is
the \hei\ constraint.}
\label{pl_h}
\end{figure}

\begin{figure}
\centerline{\psfig{file=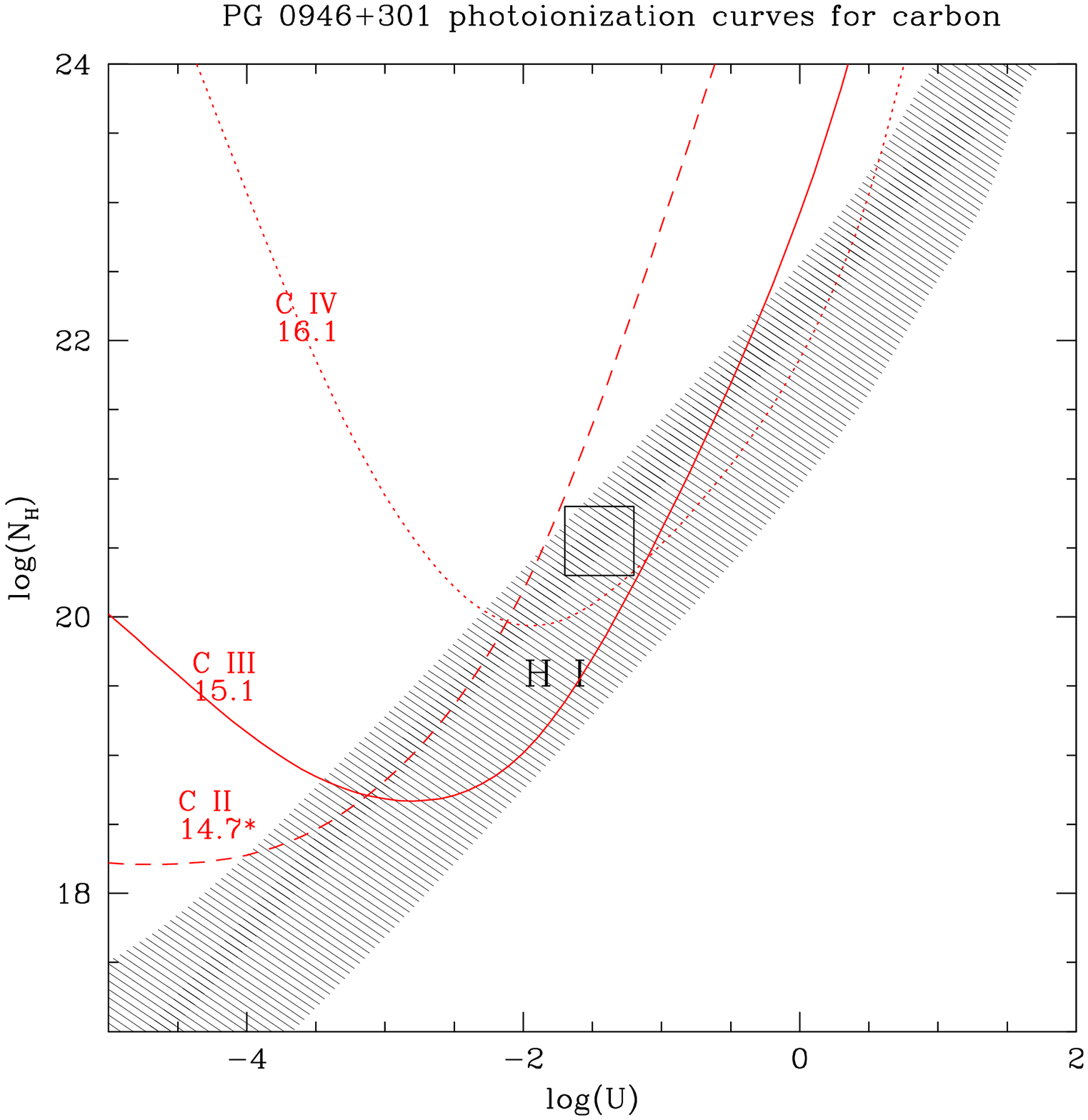,height=10.0cm,width=10cm} 
\psfig{file=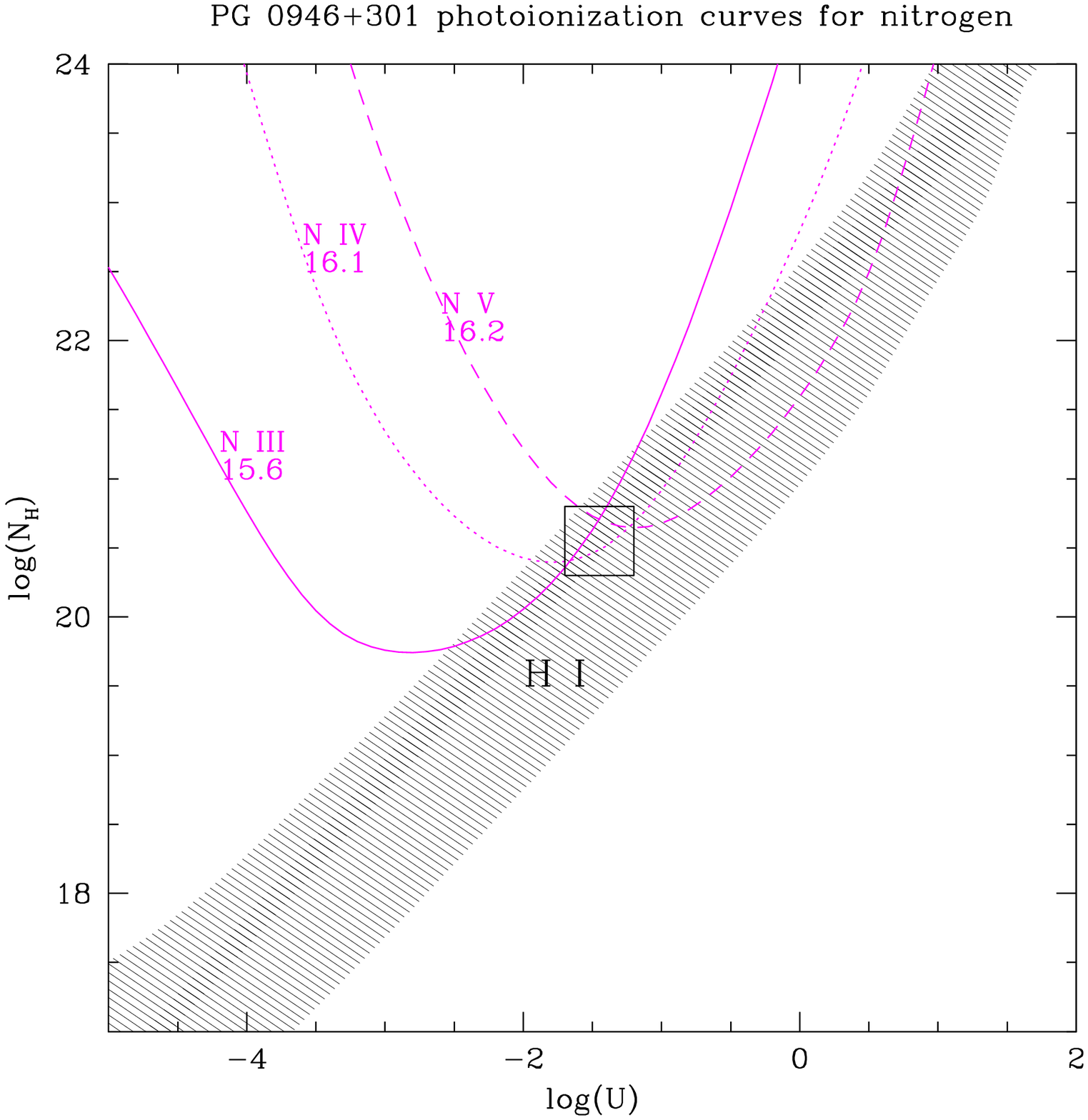,height=10.0cm,width=10cm}}
\centerline{\psfig{file=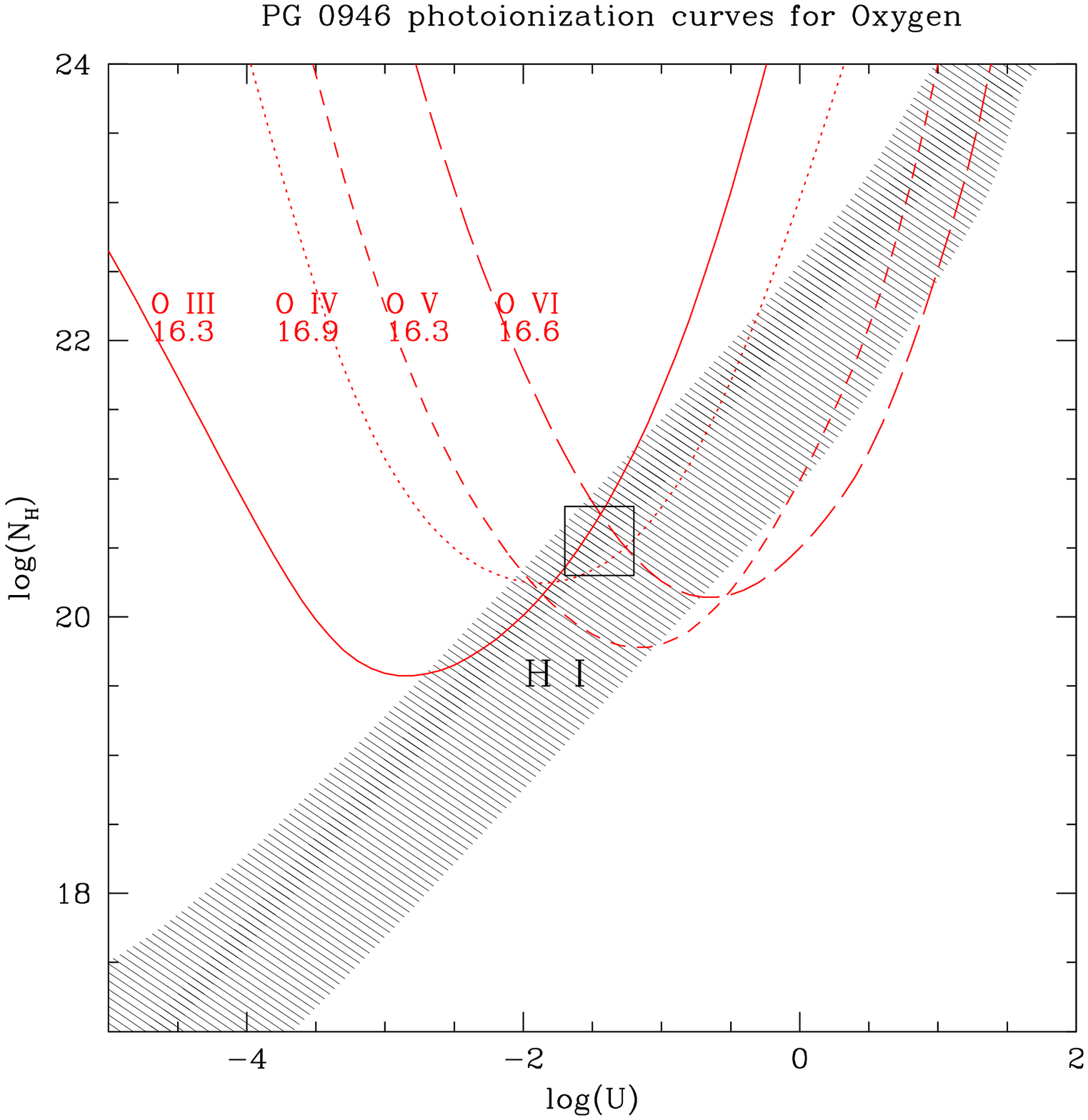,height=10.0cm,width=10cm}
\psfig{file=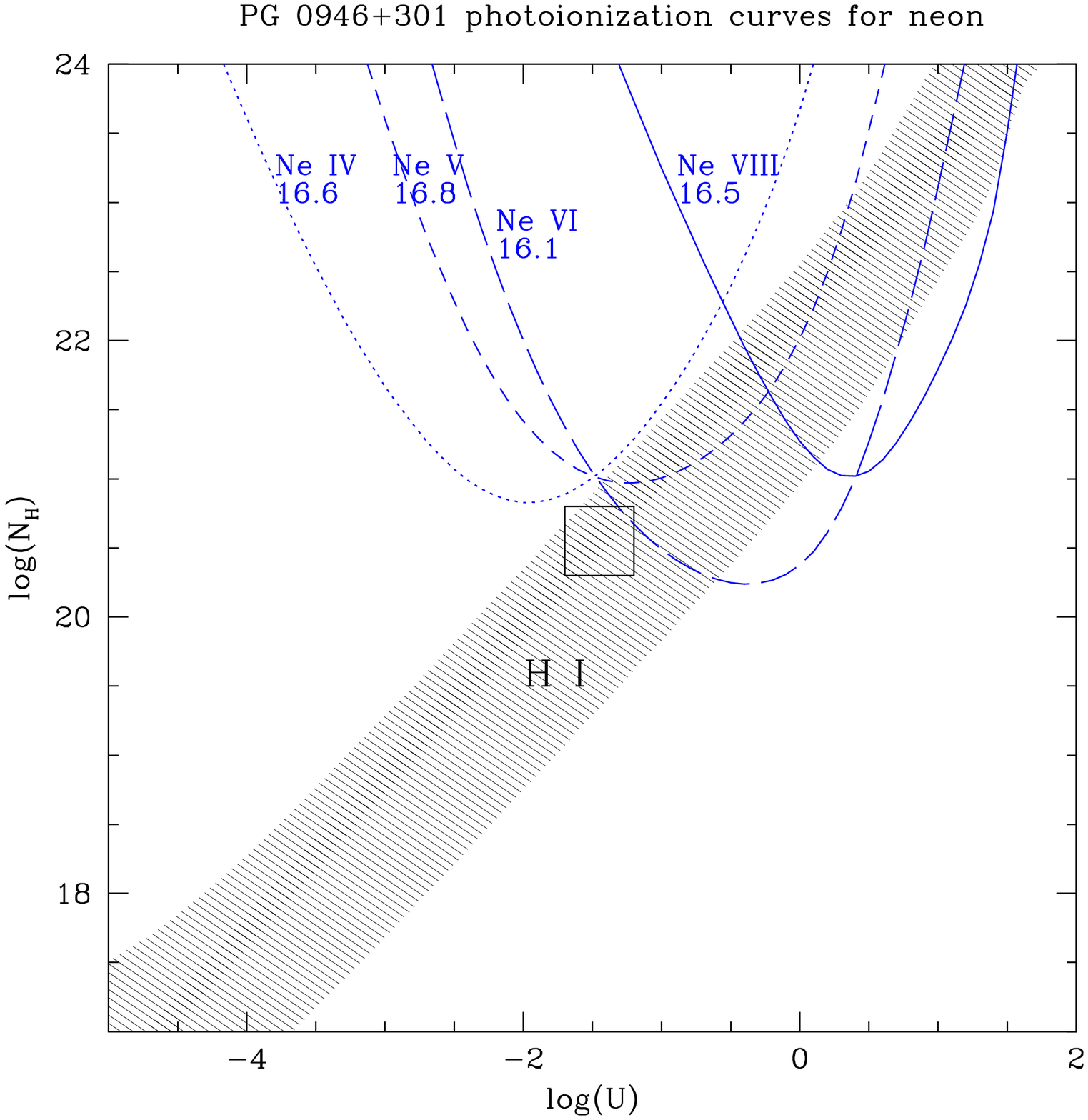,height=10.0cm,width=10cm}}
\caption{Metal-ion constraints for the 
modified $\alpha=-2$ power-law spectrum, an~*
following the $\log(N_{ion})$ value denotes an upper limit (excluding the
parameter area above the curve), otherwise the constraint is a lower
limit (excluding the parameter area below the curve).  }
\label{pl_metals}
\end{figure}

\begin{figure}
\centerline{\psfig{file=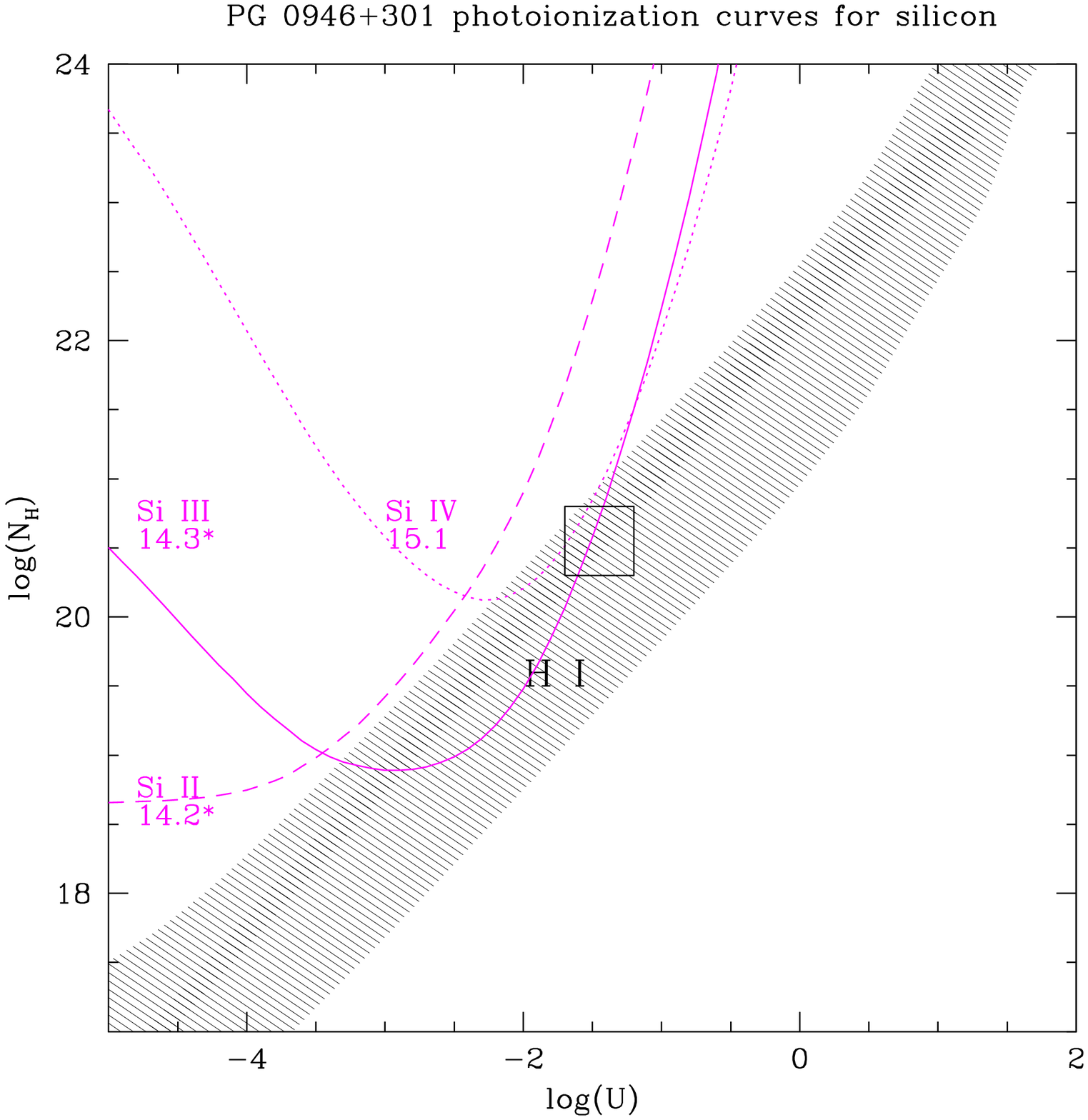,height=10.0cm,width=10cm}
\psfig{file=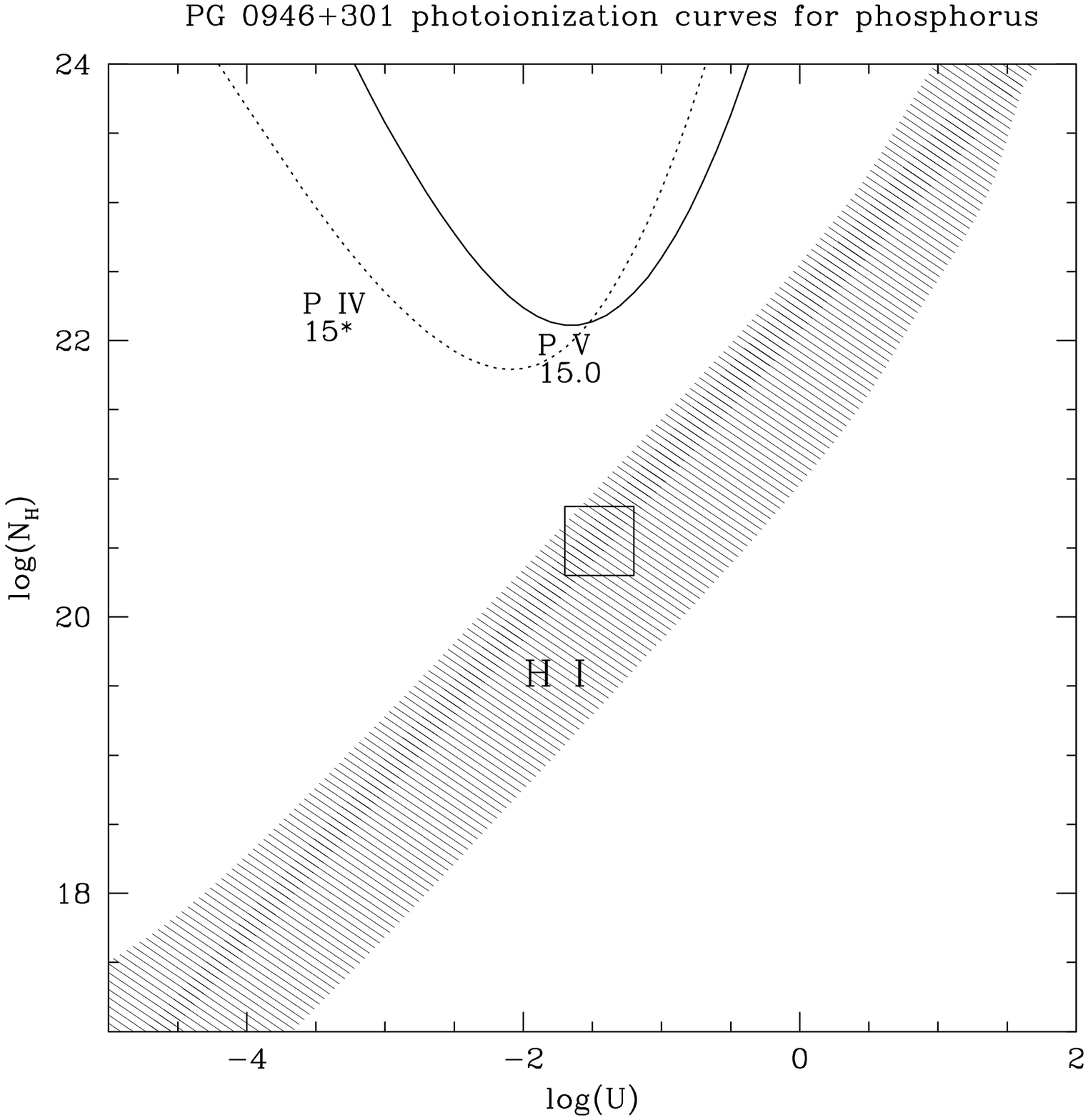,height=10.0cm,width=10cm}}
\centerline{\psfig{file=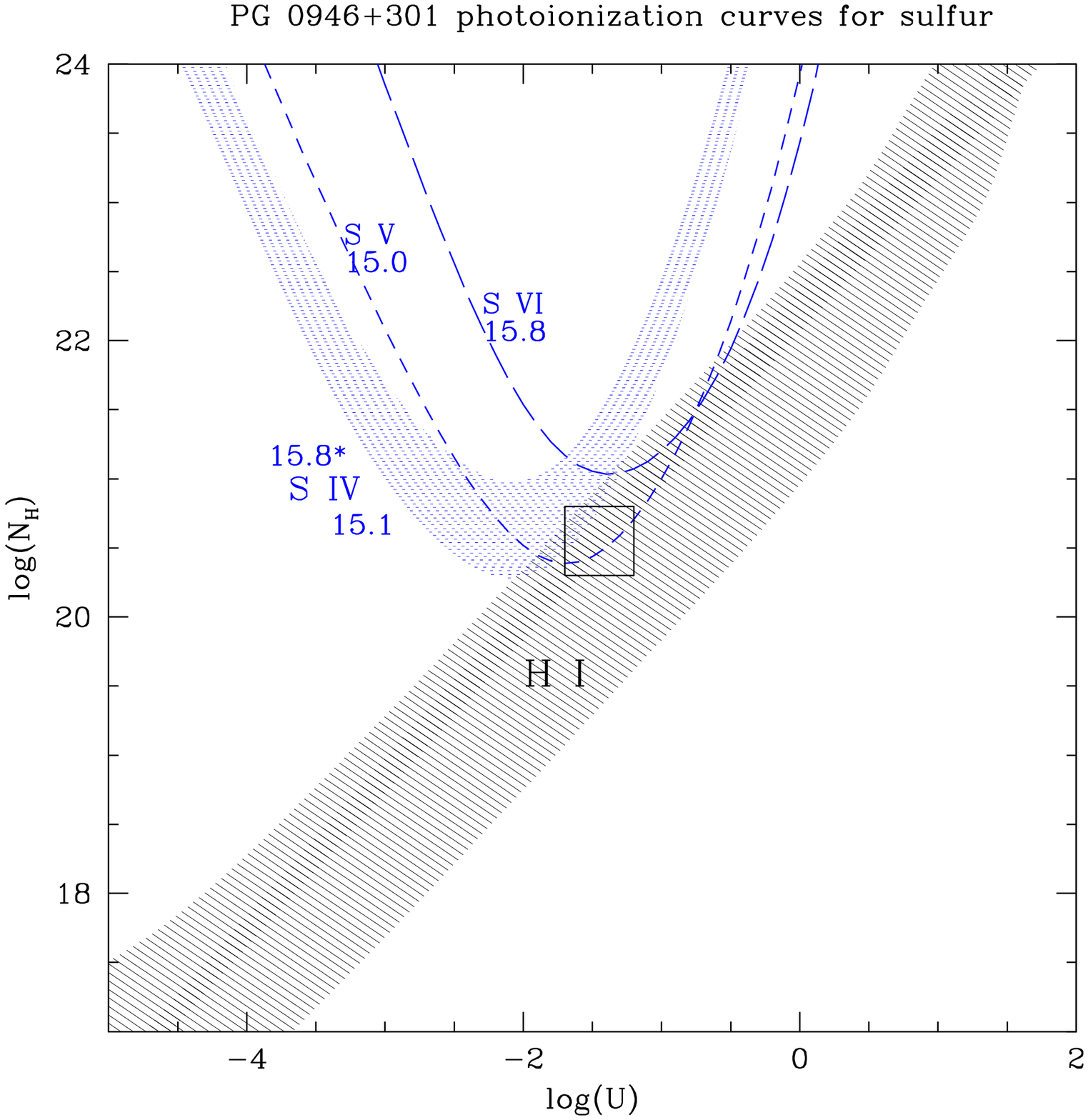,height=10.0cm,width=10cm}
\psfig{file=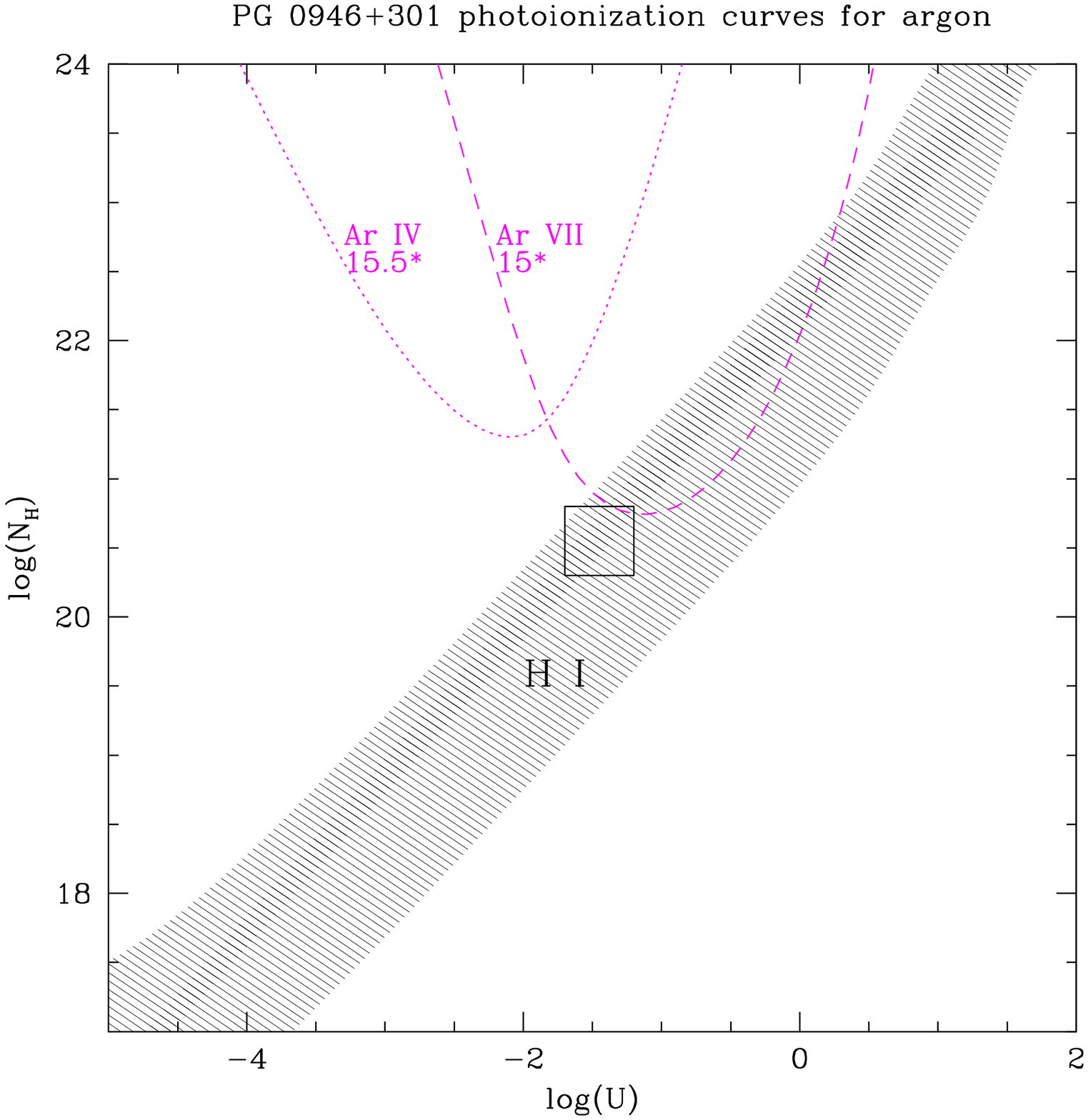,height=10.0cm,width=10cm}}
Fig. 3 continued 
\label{}
\end{figure}


\begin{figure}
\centerline{\psfig{file= 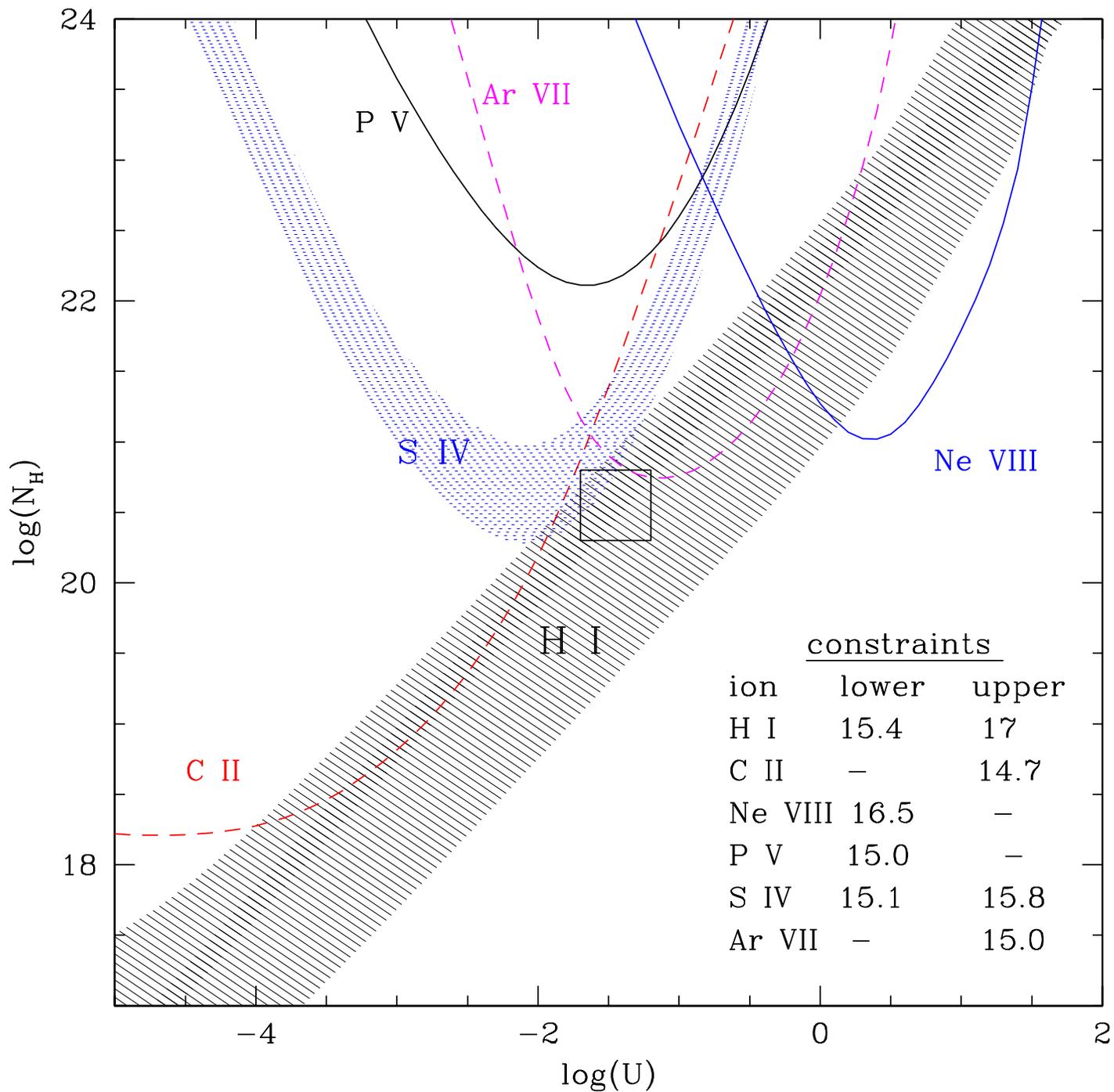,height=20.0cm,width=20.0cm}}
\caption{Some of the more important ionization constraints from several elements
(identical to Fig.~7 in the main paper).}
\label{pl_fig7}
\end{figure}


\begin{figure}
\centerline{\psfig{file=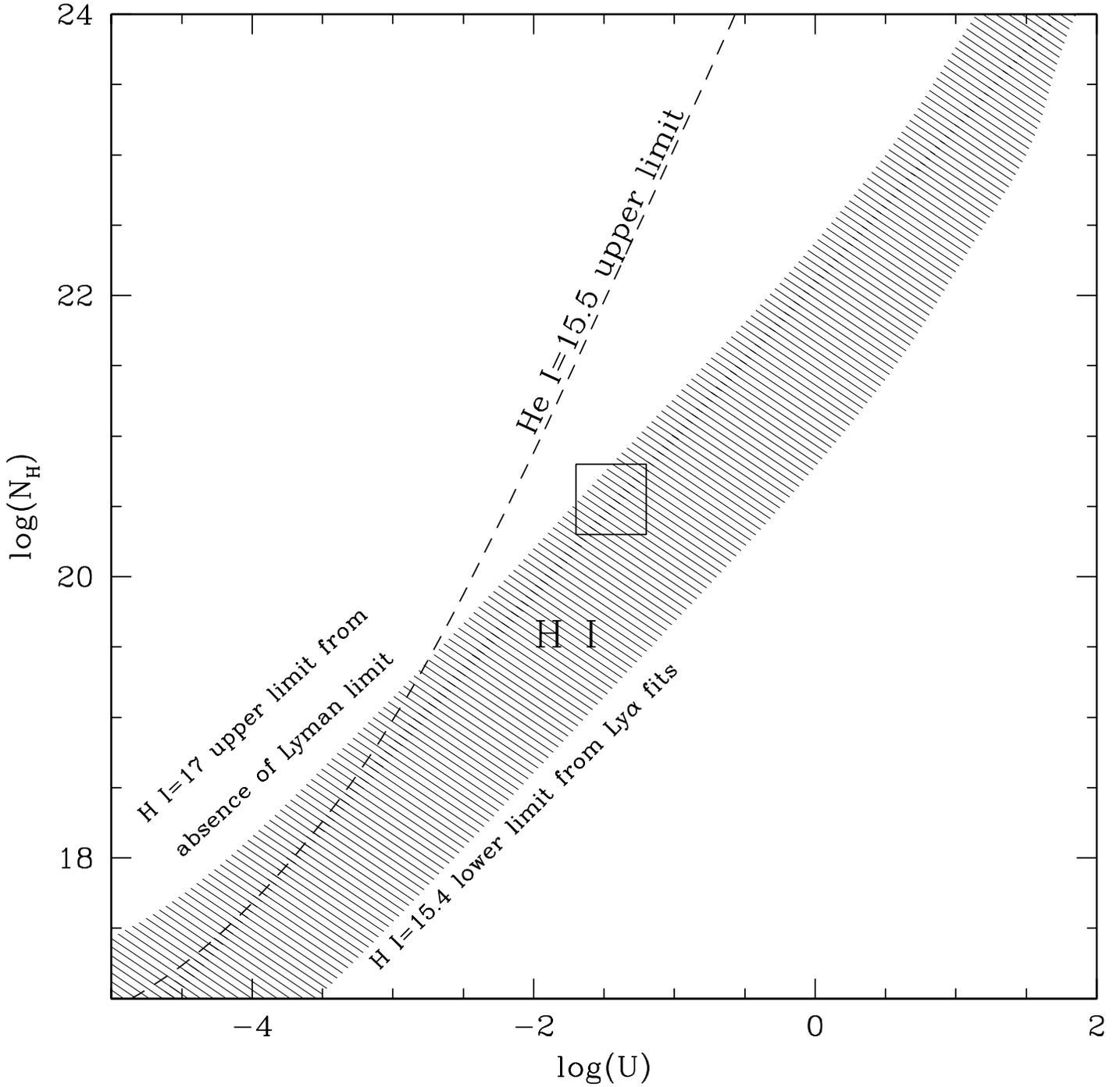,height=20.0cm,width=20.0cm}}
\caption{The all-important hydrogen constraints and their origin for
the Mathews--Ferland spectrum.  Also shown is the \hei\ constraint.}
\label{mf_h}
\end{figure}

\begin{figure}
\centerline{\psfig{file=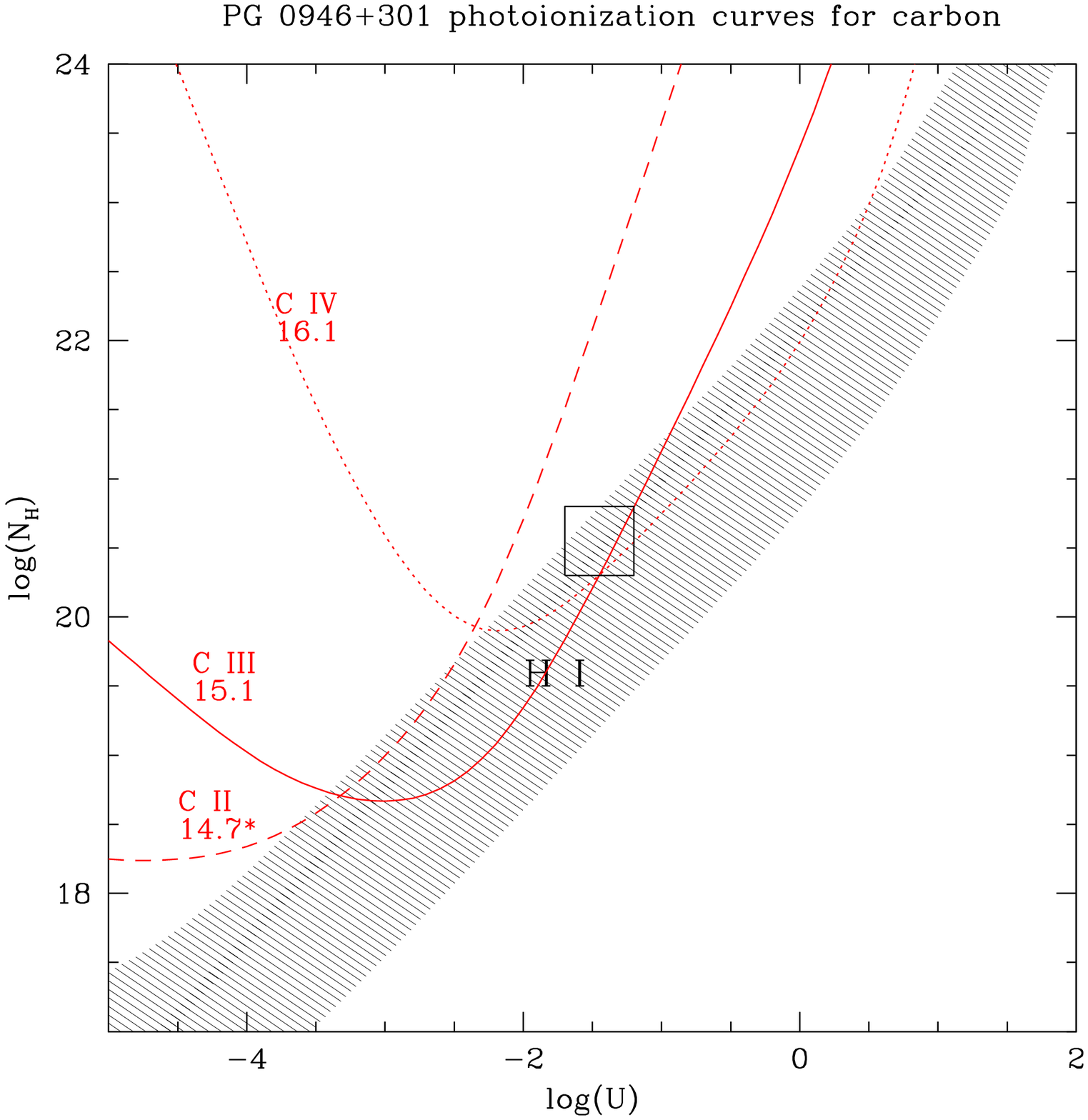,height=10.0cm,width=10cm} 
\psfig{file=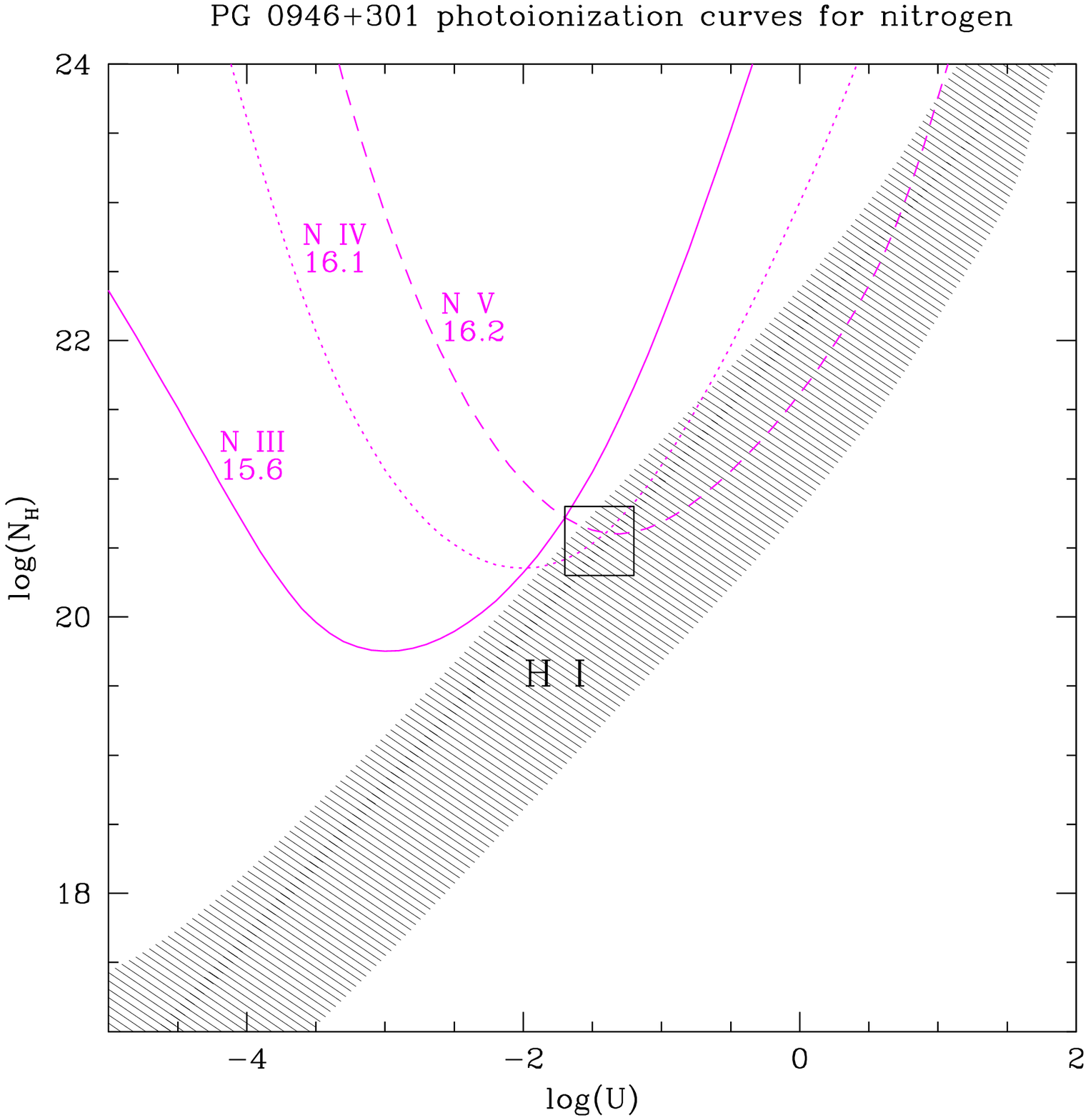,height=10.0cm,width=10cm}}
\centerline{\psfig{file=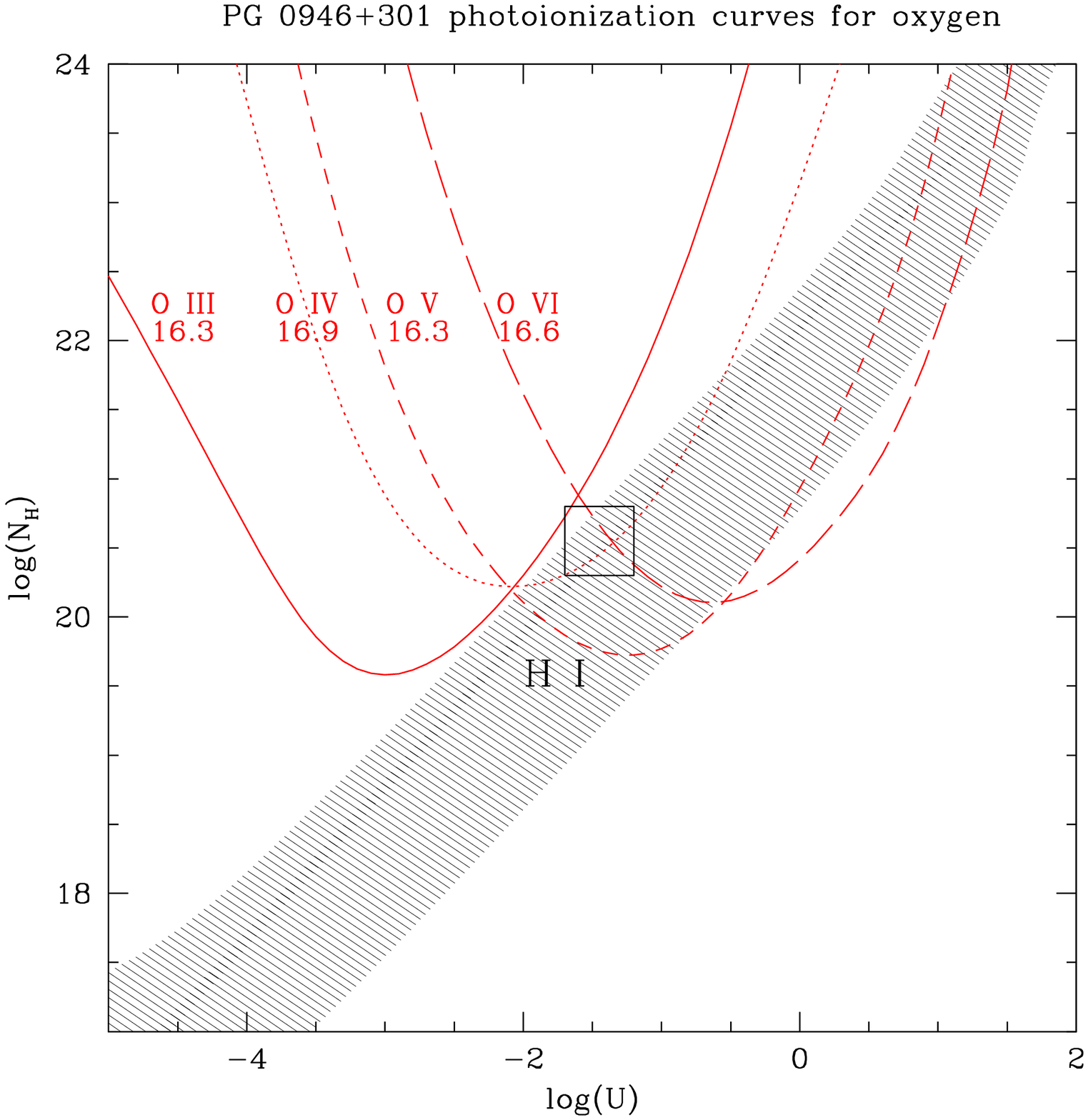,height=10.0cm,width=10cm}
\psfig{file=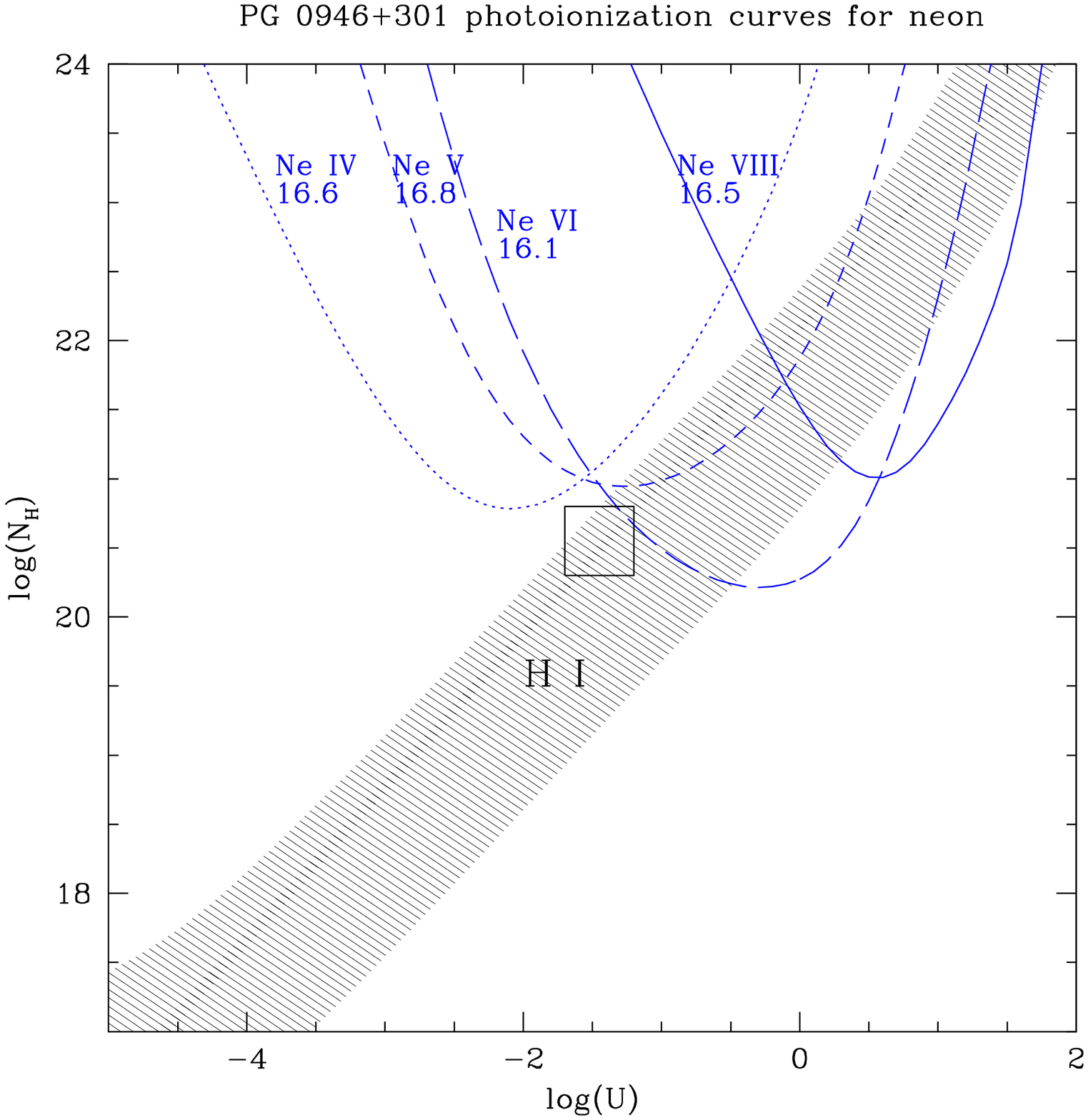,height=10.0cm,width=10cm}}
\caption{Metal-ion constraints for the Mathews--Ferland spectrum, an~*
following the $\log(N_{ion})$ value denotes an upper limit (excluding the
parameter area above the curve), otherwise the constraint is a lower
limit (excluding the parameter area below the curve).  }
\label{mf_metals}
\end{figure}

\begin{figure}
\centerline{\psfig{file=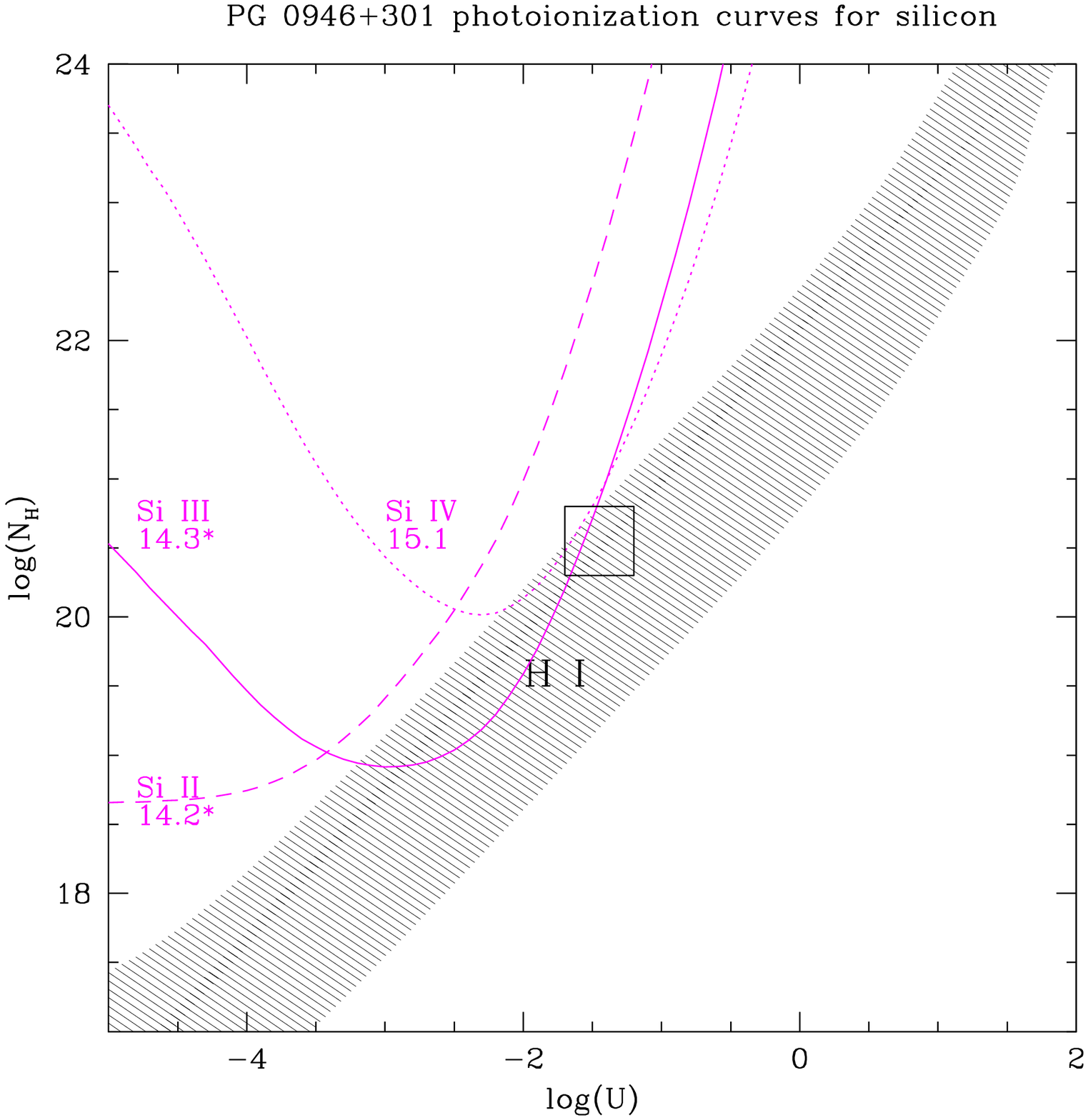,height=10.0cm,width=10cm}
\psfig{file=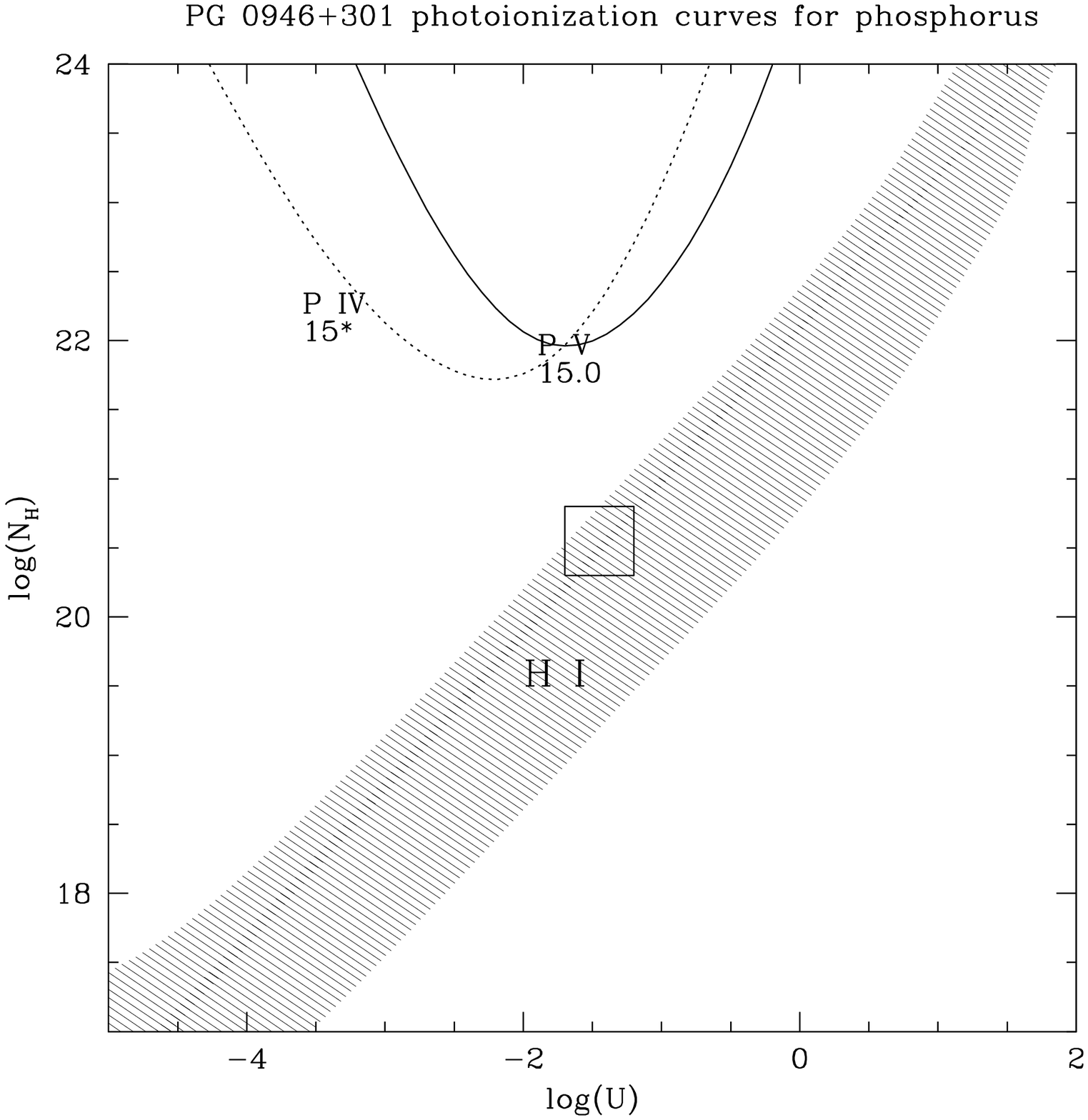,height=10.0cm,width=10cm}}
\centerline{\psfig{file=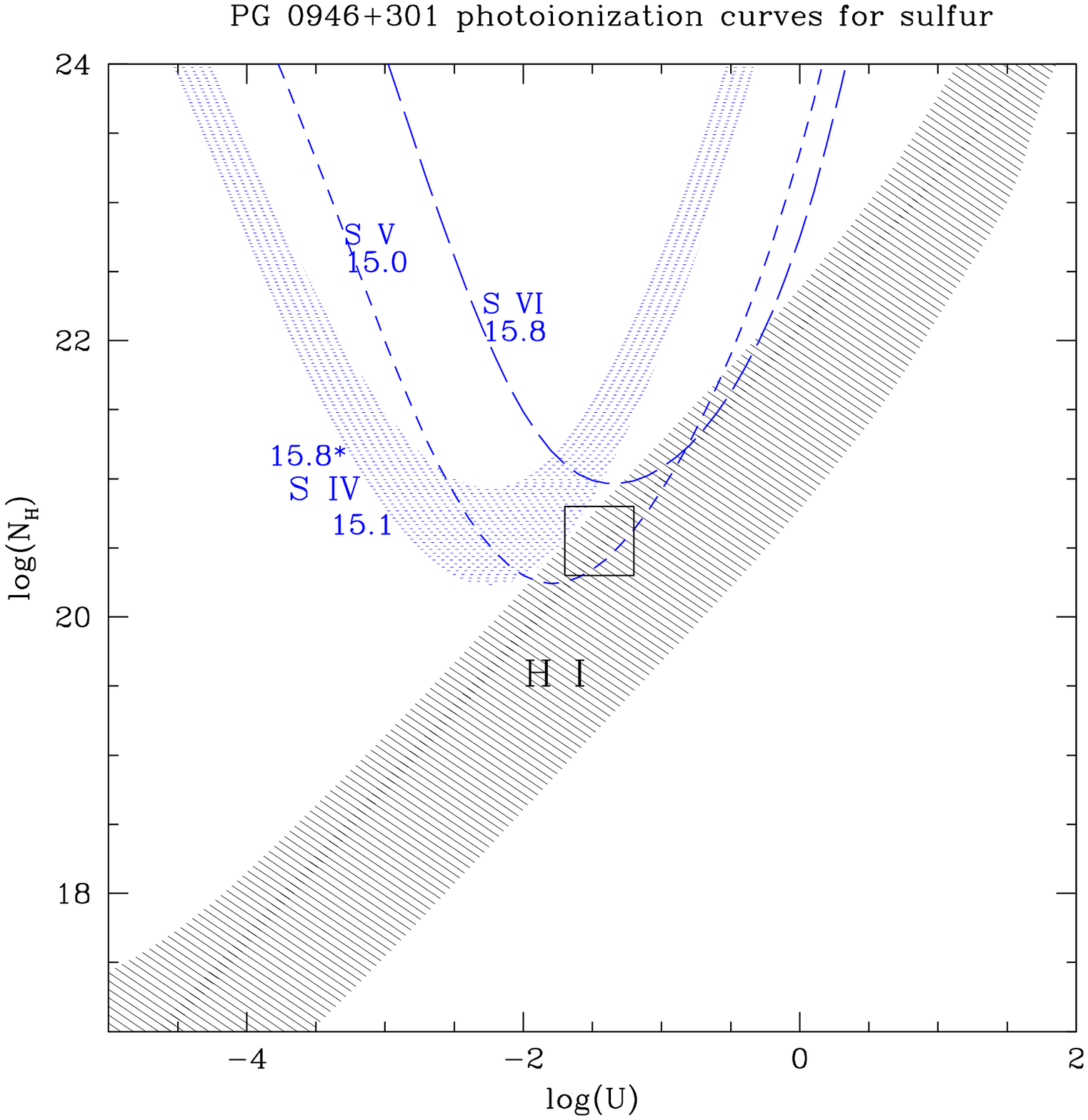,height=10.0cm,width=10cm}
\psfig{file=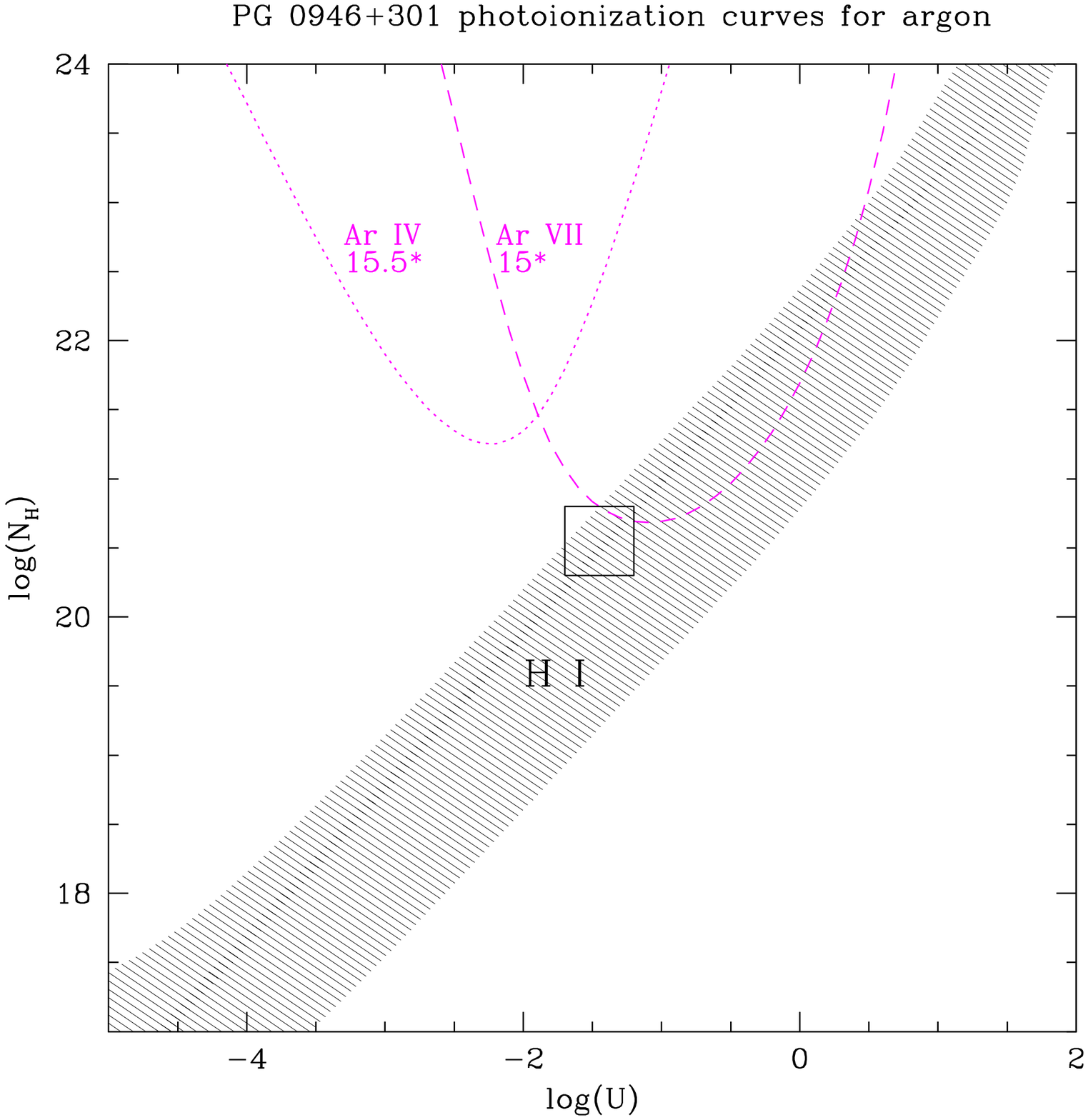,height=10.0cm,width=10cm}}
Fig. 6 continued 
\end{figure}


\begin{figure}
\centerline{\psfig{file=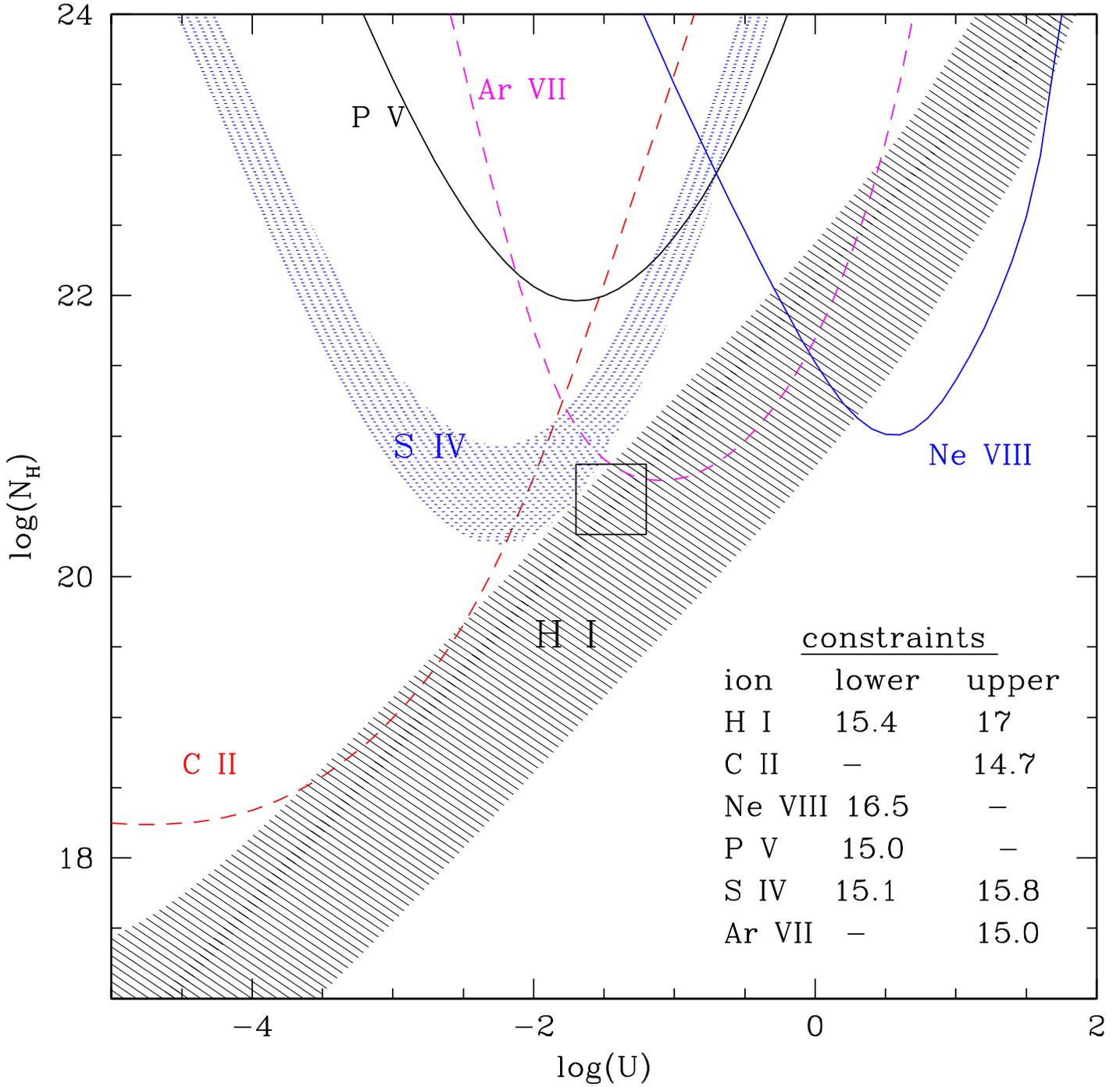,height=20.0cm,width=20.0cm}}
\caption{Some of the more important ionization constraints from several elements
for the Mathews--Ferland spectrum. } 
\label{mf_fig7}
\end{figure}

\end{document}